\documentclass[12pt]{article}

\usepackage{graphicx}  
\graphicspath{{figs/}}
\usepackage{dcolumn}   
\usepackage{bm}        
\usepackage{amssymb}   
\usepackage{amsmath}   
\usepackage{amsthm}    
\usepackage{xcolor}    
\usepackage{colortbl} 
\usepackage{tikz}      
\usepackage{ifthen}    
\usepackage{multirow}  
\usepackage{tabu}      
\usepackage{soul}      
\usepackage{subfigure}
\usepackage{xspace}
\usepackage{float}
\usepackage{adjustbox}
\usepackage{geometry}
\usepackage{etoolbox}
\usepackage{xfp}
\usepackage{authblk}
\usepackage[unicode=true,
  linktocpage,
  linkbordercolor={0.5 0.5 1},
  citebordercolor={0.5 1 0.5},
  linkcolor=blue]{hyperref}
\usepackage[numbers]{natbib}

\newcommand{\highTc}{high-$T_{\rm c}$\xspace}
\newcommand{\Tc}{$T_{\rm c}$\xspace}
\newcommand{\NdLSCO}{La$_{\rm 1.6-x}$Nd$_{\rm 0.4}$Sr$_x$CuO$_4$\xspace}

\newcommand{\ps}{\ensuremath{p^{\star}}\xspace}
\newcommand{\Ts}{$T^{\rm \star}$\xspace}
\newcommand{\degr}[1]{\ensuremath{#1^{\circ}}\xspace}

\definecolor{editblue}{rgb}{.4,.4,1}
\definecolor{mylightbeige}{HTML}{F5E6CE}

\newcounter{para}

\begin{document}

\title{Fermi surface transformation at the pseudogap critical point of a cuprate superconductor}

\author[1]{Yawen Fang\footnote{Contributed equally.}}
\newcommand\CoAuthorMark{\footnotemark[\arabic{footnote}]}
\author[1,2,3]{Ga\"el Grissonnanche\protect\CoAuthorMark}
\author[3,4]{Ana\"elle Legros}
\author[3]{Simon Verret}
\author[3]{Francis Lalibert\'e}
\author[3]{Cl\'ement Collignon}
\author[3]{Amirreza Ataei}
\author[3]{Maxime Dion}
\author[5]{Jianshi Zhou}
\author[6]{David Graf}
\author[7]{M.~J.~ Lawler}
\author[8]{Paul A. Goddard}
\author[3,9]{Louis Taillefer}
\author[1,9]{B.~J.~Ramshaw \thanks{bradramshaw@cornell.edu}}

\affil[1]{Laboratory of Atomic and Solid State Physics, Cornell University, Ithaca, NY, USA}
\affil[2]{Kavli Institute at Cornell for Nanoscale Science, Ithaca, NY, USA}
\affil[3]{D\'epartement de physique, Institut quantique, and RQMP, Universit\'e de Sherbrooke, Sherbrooke, Qu\'ebec, Canada}
\affil[4]{SPEC, CEA, CNRS-UMR 3680, Universit\'e Paris-Saclay, Gif-sur-Yvette, France }
\affil[5]{Materials Science and Engineering Program, Department of Mechanical Engineering, University of Texas at Austin, Austin, TX, USA}
\affil[6]{National High Magnetic Field Laboratory, FL, USA}
\affil[7]{Department of Physics, Applied Physics and Astronomy, Binghamton University, Binghamton, New York, USA}
\affil[8]{Department of Physics, University of Warwick, Coventry, UK}
\affil[9]{Canadian Institute for Advanced Research, Toronto, Ontario, Canada}

\newpage
\date{}
\maketitle

\newpage
\section*{Abstract}

\textbf{
The nature of the pseudogap phase remains a major puzzle in our understanding of cuprate high-temperature superconductivity. Whether or not this metallic phase is defined by any of the reported broken symmetries, the topology of its Fermi surface remains a fundamental open question. Here we use angle-dependent magnetoresistance (ADMR) to measure the Fermi surface of the cuprate \NdLSCO. Outside of the pseudogap phase we can fit the ADMR data and extract a Fermi surface geometry that is in excellent agreement with angle-resolved photoemission data. Within the pseudogap regime the ADMR is qualitatively different, revealing a transformation of the Fermi surface. We can rule out changes in the quasiparticle lifetime as the sole cause of this transformation. We find that our data are most consistent with a pseudogap Fermi surface that consists of small, nodal hole pockets, thereby accounting for the drop in carrier density across the pseudogap transition found in several cuprates.}

\section*{Introduction}
A long-standing mystery of the \highTc cuprate superconductors is the `pseudogap phase' \cite{keimer2015quantum}---a correlated electronic state whose key characteristic is a loss of coherent quasiparticles below an onset temperature \Ts and below a critical doping \ps. This loss of quasiparticles is reminiscent of the superconducting gap that opens at the transition temperature \Tc (hence the name `pseudogap'), suggesting that the pseudogap phase and superconductivity are related. Characterizing what remains of the coherent Fermi surface (FS) inside the pseudogap phase is, therefore, a critical step toward understanding how this peculiar metallic state gives rise to, or is compatible with, high-temperature superconductivity.

\sloppypar Heavily overdoped cuprates are good metals with a well-defined FS. Tl$_2$Ba$_2$CuO$_{6 + \delta}$ (Tl2201) has been measured extensively in this doping regime and three independent experiments agree on the geometry of the FS: angle-dependent magnetoresistance (ADMR)~\cite{hussey2003coherent}, angle-resolved photoemission spectroscopy (ARPES)~\cite{plate2005fermi}, and quantum oscillations~\cite{vignolle2008quantum}. Other cuprates, such as La$_{2-x}$Sr$_x$CuO$_4$ and Bi$_2$Sr$_2$CaCu$_2$O$_{8+\delta}$, show similar agreement between the measured FS and band structure calculations for $p>\ps$\cite{damascelli2003angle}. As the doping is lowered toward \ps the Fermi surface measured by ARPES remains well-defined but the electrical resistivity becomes progressively more anomalous, becoming perfectly linear-in-temperature at \ps \cite{cooper2009anomalous,daou2009linear}. Whether a $T$-linear scattering rate alone can account for this anomalous resistivity has been the subject of much debate: we have addressed this topic in a recent study \cite{grissonnanche2020measurement}.

Cuprates enter the pseudogap phase below \ps. While this phase is also metallic, its FS---in the limit $T\rightarrow 0$ and in the absence of superconductivity---remains unknown. ARPES measurements performed above \Tc~and below \Ts find discontinuous segments known as ``Fermi arcs'' \cite{damascelli2003angle}, which defy the conventional definition of a closed FS. Quantum oscillations, on the other hand, reveal a small, closed, electron-like FS (``electron pocket'') \cite{doiron2007quantum}. This pocket, however, appears only in the presence of charge density wave (CDW) order \cite{ramshaw2015quasiparticle}, and CDW order is not always observed over the same range of dopings as the pseudogap phase itself. For example, while CDW order extends up to \ps in HgBa$_2$CuO$_{4+x}$ \cite{chan2020extent}, it terminates before \ps at $p \approx 0.16$ in YBa$_2$Cu$_3$O$_{6+x}$ \cite{blanco2014resonant} and at $p \approx 0.18$ in La$_{1.6-x}$Nd$_{0.4}$Sr$_x$CuO$_4$ \cite{gupta2020vanishing} (the compound we study here.) Spin density wave (SDW) order has also been found below \ps in several cuprates \cite{tranquada1997coexistence,ma2021parallel, franchet2020hidden}. Recent neutron diffraction measurements have even found indications of SDW order at $p = 0.24$, in zero magnetic field and at $T = 13$~K \cite{ma2021parallel} (see \autoref{fig:phasedat}). While SDW order is known to reconstruct part of the Fermi surface at much lower doping \cite{kunisada2020observation}, our sample at $p = 0.24$ shows perfectly linear resistivity down to $2$~K at $B=35$~T, without an upturn at $T = 13$~K that would be characteristic of SDW order \cite{bourgeois2019link} (see Extended Data Figure 1). This suggests that either there are differences between samples grown by different groups or that a magnetic field suppresses the SDW order at $p=0.24$.

A crucial question therefore remains: what is the Fermi surface of cuprates immediately below \ps in the absence of superconductivity or CDW order? There are two possibilities: i)~the FS is the same above and below \ps, but the quasiparticles become incoherent below \ps due to scattering or other correlation effects; ii)~the FS below \ps is different from the FS above \ps. Demonstration of the latter scenario would imply either that translational symmetry is broken (on some appropriate length scale) in the pseudogap phase or that it is a phase with topological order \cite{scheurer2018topological}.

\section*{Experiment}

To investigate the possibility of Fermi surface reconstruction below \ps, we turn to the cuprate La$_{1.6-x}$Nd$_{0.4}$Sr$_x$CuO$_4$ (Nd-LSCO). The critical doping \ps$ = 0.23$ that marks the onset of the pseudogap phase in Nd-LSCO has been well-characterized by transport \cite{daou2009linear,collignon2017fermi}, specific heat \cite{michon2019thermodynamic} and ARPES \cite{matt2015electron}. At $p=0.20$, a gap opens along the “anti-nodal” directions of the Brillouin zone ($\phi = \degr{0}$, \degr{90}, \degr{180}, and \degr{270}) upon cooling below \Ts$ = 75$ K, followed by an upturn in the resistivity; at $p=0.24$, ARPES detects no anti-nodal gap and the resistivity remains perfectly linear down to the lowest measured temperature. Note that the highest doping where X-ray scattering detects CDW order in Nd-LSCO is $p=0.17$ \cite{gupta2020vanishing}. As with other cuprates \cite{adachi2001crystal,leboeuf2007electron}, the onset of CDW order in Nd-LSCO coincides with a downturn of the Hall coefficient toward negative values \cite{noda1999evidence}. At $p=0.20$ and above, the Hall coefficient remains positive at all temperatures and magnetic fields \cite{collignon2017fermi}. This suggests that the quasiparticles responsible for transport (and hence ADMR) do not feel the influence of any remnant CDW order at the dopings where we perform our measurements, in agreement with the absence of any CDW modulations detected by X-ray diffraction and the Seebeck coefficient at $p = 0.18$ and above \cite{michon:2018wiedemann, gupta2020vanishing}.

To determine whether the FS is transformed across \ps we measure variations in the $c$-axis resistivity $\rho_{\rm zz}$ of Nd-LSCO at $p=0.21$ and $p=0.24$ as a function of the polar ($\theta$) and azimuthal ($\phi$) angles between the sample and an external magnetic field $\bm{B}$ (see \autoref{fig:phasedat}b, c, d)---a technique known as angle-dependent magnetoresistance (ADMR). These variations are determined by the three-dimensional geometry of the Fermi surface and the momentum-dependence of the scattering rate. The basic premise of ADMR is that the velocities of charge-carrying quasiparticles are modified by the Lorentz force in a magnetic field. Within the standard relaxation-time approximation, $\rho_{\rm zz}$ is given by Chambers' solution to the Boltzmann transport equation:
\begin{equation}
1/\rho_{zz}  = \frac{e^2}{4 \pi^3}\!\oint \!d^2\bm{k} ~\mathcal{D}\!\left(\bm{k}\right) v_z\!\left[\bm{k}\!\left(t=0\right)\right]\int^{0}_{-\infty}\!v_z\!\left[\bm{k}\!\left(t\right)\right]e^{t/\tau}dt,
\label{eq:chambz}
\end{equation}
where $\oint \!d^2\bm{k}$ is an integral over the Fermi surface, $\mathcal{D}\!\left(\bm{k}\right)$ is the density of states at point $\bm{k}$, $v_z$ is the component of the Fermi velocity in the $z$ direction, and  $\int^{0}_{-\infty}\!v_z\!\left[\bm{k}\!\left(t\right)\right]e^{t/\tau}dt$ is an integral of $v_z$ weighted by the probability that a quasiparticle with lifetime $\tau$ scatters after time $t$~\cite{Chambers1952kinetic}. The magnetic field $\bm{B}$ enters through the Lorentz force, $\hbar \frac{d\bm{k}}{dt} = e~ \bm{v}\!\times\!\bm{B}$, which induces the quasiparticles into cyclotron motion around the Fermi surface (see \autoref{fig:overdoped}c).

For a quasi-two dimensional Fermi surface with simple, sinusoidal warping, $\rho_{zz}$ can be calculated analytically from \autoref{eq:chambz} in the limit where $\tau$ is long \cite{yamaji1989angle}. The exact calculation contains special ``Yamaji'' angles where all cyclotron orbits have the same cross-sectional area perpendicular to the magnetic field and where the $v_z$ component of the Fermi velocity averages to zero around each orbit. This  cancellation of $v_z$ results in maxima in the $c$-axis resistivity at these angles. The Yamaji angles are determined by the geometry of the FS, and therefore by measuring the angular positions of the resistivity maxima one can construct the FS geometry. For more complex FS geometries, \autoref{eq:chambz} must be calculated numerically but the intuition still holds---at certain angles the resistivity is maximized because $v_z$ is more effectively averaged toward zero (see Supplementary Figure S1 for more information).

To reconstruct the FS geometry from the ADMR data we start with a tight-binding model $\epsilon(\bm{k})$ that respects the geometry of the transfer integrals of the material, define the Fermi velocity through $\bm{v} = \frac{1}{\hbar}\bm{\nabla_{\rm k}}\epsilon$, and then tune the tight-binding parameters until the calculated $\rho_{zz}$ matches the measured data. In addition to the FS geometry, ADMR is sensitive to the momentum dependence of the quasiparticle scattering. This is captured in \autoref{eq:chambz} by introducing $\tau(\bm{k})$---the full expression for $\rho_{zz}$ in this case is given in the Methods. We separate the scattering rate into isotropic and anisotropic components, $1 / \tau(\bm{k}) = 1/\tau_{\rm iso}+1/\tau_{\rm aniso}(\bm{k})$. These two components can have distinct temperature dependences as demonstrated in Tl-2201 \cite{abdel2006anisotropic}. The approach of using \autoref{eq:chambz} to determine Fermi surfaces has been particularly successful in 2D metals such as organic conductors \cite{singleton2000studies} and Sr$_2$RuO$_4$ \cite{bergemann2003quasi}. In cuprates, ADMR has been measured in the overdoped regime ($p>\ps$) \cite{hussey2003coherent,grissonnanche2020measurement}, in the underdoped regime with CDW order ($p \approx 0.1 \ll \ps$)\cite{ramshaw2017broken}, and in electron-doped materials \cite{kartsovnik2011fermi}, but never in the pseudogap phase in the absence of CDW order.

\section*{Doping $\mathbf{p=0.24 > p^*}$}

\autoref{fig:phasedat}d shows the ADMR of Nd-LSCO at $p=0.24$, at $T=25$~K and $B = 45$~T. We fit the data using a single-band tight-binding model that is commonly used for cuprates with a body centred tetragonal unit-cell (see Methods for the full model). We then perform a global optimization over the tight binding and scattering rate parameters using a genetic algorithm, placing loose bounds on the parameters around values determined by previous ARPES measurements \cite{matt2015electron,horio2018three}. The right panel of \autoref{fig:overdoped}b shows the results of this optimization: key features reproduced by the fit include the position of the maximum near $\theta = \degr{40}$, the onset of $\phi$-dependence beyond $\theta = \degr{40}$, and the $\phi$-dependent peak/dip near $\theta = \degr{90}$. The best-fit tight-binding parameters are in good agreement with those determined by ARPES (see Extended Data Table 1), demonstrating excellent consistency between the two techniques.

The peak in $\rho_{zz}$ near $\theta = \degr{40}$, which is captured well by the fit, can also be checked against the intuitive picture of ADMR described earlier: the position of this peak should be related to the length of the Fermi wavevector, $k_{\rm F}$. For a Fermi surface with the simplest sinusoidal dispersion along $k_z$, an analytic calculation of \autoref{eq:chambz} shows that the ADMR changes with angle as $\rho_{zz} \propto 1/\left(J_0\left(c k_F \tan\theta\right)\right)^2$, where $c$ is the inter-layer lattice constant, $k_{\rm F}$ is the Fermi wavevector, and $J_0$ is the $0^{\mathrm{th}}$ Bessel function of the first kind. While the analytic expression for $\rho_{zz}$ is not exact for the particular form of interlayer hopping found in Nd-LSCO, in Supplementary Figure S1 we show that the maxima in the resistivity coincide with the angles where $v_z$ is best averaged to zero. Our analysis including the proper interlayer hopping shows that the peak in $\rho_{zz}$ near $\theta = \degr{40}$ suggests that $k_F \approx 7$~nm$^{-1}$ along the zone diagonal, which is very close to the FS shown in \autoref{fig:overdoped}c. This suggests that the ADMR at $p=0.24$ exhibits features consistent with a large, unreconstructed Fermi surface, as also observed by ARPES.

In addition to the Fermi surface geometry, ADMR is sensitive to the momentum-dependent quasiparticle scattering rate. We find that the $p=0.24$ data is best described by a highly anisotropic scattering rate that is largest near the anti-nodal regions of the Brillouin zone and smallest near the nodal regions. More details of the scattering rate model and its temperature dependence can be found in \citet{grissonnanche2020measurement}.

\section*{Doping $\mathbf{p=0.21 < p^*}$}

We now turn to Nd-LSCO $p=0.21$, below \ps and inside the pseudogap phase, where ARPES finds discontinuous segments of FS known as `arcs' \cite{damascelli2003angle}. Upon comparison of \autoref{fig:phasedat}c and \autoref{fig:phasedat}d, it is immediately apparent that the structure of the ADMR changes qualitatively upon entering the pseudogap phase. In particular, the resistivity peak near $\theta = \degr{40}$ has disappeared at $p=0.21$. The qualitative differences in the data arise either from a change in the FS geometry, or from a large increase in the scattering rate for the anti-nodal quasiparticles (e.g. the generation of Fermi arcs).

We test several different scenarios to understand the change in the ADMR across \ps. These scenarios can be divided into two classes: those that change only the quasiparticle scattering rate, and those that reconstruct the Fermi surface. First, we use the same FS model and scattering rate that fit the ADMR at $p=0.24$ and simply adjust the chemical potential to decrease the hole concentration to $p=0.21$. The simulated data for this model are shown in \autoref{fig:underdoped}a. Instead of describing the data for $p=0.21$, however, this simulation appears close to that for $p=0.24$. This is to be expected: only the FS near the anti-nodal region changes appreciably upon lowering the doping, and the ADMR is less sensitive to this region due to its high scattering rate. Therefore, something beyond a simple change in the chemical potential must occur when crossing \ps.

Next we test three other scattering rate models (on the large, unreconstructed FS): the same model used at $p = 0.24$ but now with the scattering rate parameters allowed to vary (Extended Data Figure 2c); isotropic scattering around the entire FS (Extended Data Figure 2b); and a model of `Fermi arcs' where the quasiparticle lifetime diminishes rapidly past the antiferromagnetic zone boundary on the FS in \autoref{fig:underdoped}c. Even after performing fits using the genetic algorithm, allowing for a broad range of band-structure and scattering-rate parameters, none of these scattering rate models on the large unreconstructed hole-like FS is able to reproduce the ADMR at $p=0.21$ (see Methods for a description of the fits and Extended Data Figure 2.) Note that the average strength of the scattering does not seem to change much as the system crosses \ps, since the magnitude of the ADMR, which is essentially governed by the magnitude of $1/\tau$, is roughly the same at $p = 0.21$ and at $p = 0.24$ (\autoref{fig:phasedat}).The inability of any of these scattering rate models to fit the ADMR at $p=0.21$ suggests that the FS must be reconstructed into a new, geometrically distinct, FS in the pseudogap phase.

To confirm a change in the FS geometry across \ps, we test two models of FS reconstruction. First, we try a small electron pocket at nodal positions in the Brillouin zone, as in \autoref{fig:underdoped}e. This FS is the result of a bi-axial charge density wave, as found in several underdoped cuprates \cite{vershinin2004local,wu2011magnetic}, and is likely the origin of the small electron pocket found in YBa$_2$Cu$_3$O$_{6+x}$ and HgBa$_2$CuO$_{4+x}$ \cite{doiron2007quantum}. This Fermi surface accounts well for the ADMR of YBa$_2$Cu$_3$O$_{6.6}$ at $p=0.11$, where there is CDW order \cite{ramshaw2017broken}. We simulate the ADMR using the method described earlier, now calculating the Fermi velocity and density of states using the reconstructed band structure, the results of which are shown in \autoref{fig:underdoped}f. Even if one allows the tight-binding and gap parameters to vary, or if one uses a $d$-wave form-factor for the CDW gap \cite{allais2014connecting}, these simulations do not agree at all with the ADMR for Nd-LSCO $p=0.21$ (see Methods and Extended Data Figure 3). This suggests that the FS transformation at $p=0.21$ is not due to the same CDW order that produces the nodal electron pocket found in other underdoped cuprates. This is consistent with the Hall and Seebeck coefficients, which remain positive at all temperatures and magnetic fields in Nd-LSCO at $p=0.21$ \cite{collignon2017fermi,collignon2021thermopower}, whereas negative (or negative-trending) Hall and Seebeck coefficients are a ubiquitous signature of charge order in the cuprates, observed in four distinct families of cuprates \cite{adachi2001crystal,leboeuf2007electron,doiron2013hall}, including Nd-LSCO at $p=0.12$ \cite{noda1999evidence}. It is also consistent with recent X-ray scattering experiments, which find no charge order at dopings greater than $x = 0.17$ in Nd-LSCO \cite{gupta2020vanishing}. 

Finally, we consider small hole pockets centred around the nodal directions of the Fermi surface, as shown in (\autoref{fig:underdoped_pipi}). Such nodal hole pockets arise in various theoretical scenarios \cite{wen1996theory,chakravarty2001hidden,rice2011phenomenological,storey2016hall,scheurer2018topological} and the Fermi arcs seen by ARPES could correspond to the front side of such pockets. In practice, we generate a FS made of four nodal hole pockets by reconstructing the large FS using antiferromagnetic order with a $Q = \left(\pi,\pi\right)$ wavevector, using the same tight-binding parameters as in the $p=0.24$ simulation. The ADMR for this FS is shown in \autoref{fig:underdoped_pipi}b. This Fermi surface reproduces all critical features of the data at $p=0.21$: the resistivity initially decreases with increasing~$\theta$; there is a minimum near $\theta = \degr{60}$; and the peak at \degr{90} is strongest along $\phi = \degr{0}$ and weakest along $\phi = \degr{45}$. Note that, despite the success of this model in reproducing the \textit{relative} change in magnetoresistance as a function of angle, the \textit{absolute} value of the resistance is off by approximately a factor of three (see Methods).

The key structures present in the reconstructed hole pockets, which are not present in the model of the arcs, are the sharp corners where the front and backsides of the hole pockets are connected: it is these corners that produce qualitatively different ADMR than is produced by the model of arcs. The gap magnitude (the strength of the potential associated with the FS reconstruction) that best reproduces the data is 5 meV, or $\approx 55$ K---this gap sets the `sharpness' of the corners on the hole pockets. Note that this gap is insufficient to remove the anti-nodal electron pockets that also result from a $Q = \left(\pi,\pi\right)$ reconstruction: we remove the electron pocket to produce agreement between the calculated and measured Hall coefficients (our data is also consistent with the inclusion of electron pockets with a much higher scattering rate than is found on the hole pockets; see Methods and Extended Data Figure 4). We find that a momentum-independent scattering rate is required to reproduce the data. This reduction in scattering-rate anisotropy between $p>\ps$ and $p<\ps$ may be due to the significant reduction in density of states anisotropy when moving from $p = 0.24$ to $p = 0.21$ (see Extended Data Figure 5, which shows a reduction of the anisotropy in the density of states from a factor of 25 at $p = 0.24$ to a factor of 2 at $p = 0.21$). Thus, the change in ADMR moving from $p = 0.24$ to $p = 0.21$ has two sources: a transformation to a Fermi surface consisting of four nodal hole pockets, and a reduction in scattering rate anisotropy.

\section*{Discussion}

Our main finding is a qualitative change in the ADMR that indicates a transformation of the FS at \ps. For $p>\ps$, excellent agreement is found between the FS measured by ADMR and the one measured by ARPES, both giving the same large, diamond-like Fermi surface \cite{grissonnanche2020measurement}. For $p<\ps$, however, the ADMR is strikingly different. This difference is not due to a simple lowering of the chemical potential through the van Hove point, nor is it solely due to a change in the scattering rate across \ps: it must therefore be due to a change in the geometry of the FS. The data below \ps are best described by a FS composed of nodal hole pockets. These nodal hole pockets can result from a $Q = \left(\pi,\pi\right)$ reconstruction. Such a reconstruction is consistent with the transition from a carrier density $n=1+p$ at $p>\ps$ to a density of $n=p$ at $p<\ps$, as revealed by the Hall coefficient \cite{badoux2016change,collignon2017fermi} (see Extended Data Figure 4 for a comparison of the measured and calculated Hall coefficients). Similar nodal hole pockets were recently detected by both quantum oscillations and ARPES in the 5-layer cuprate Ba$_2$Ca$_4$Cu$_5$O$_{10}$(F,O)$_2$ at a doping where long-range AFM order is known to exist \cite{kunisada2020observation}; the question is whether a similar reconstruction takes place in Nd-LSCO at $p = 0.21$, given that the spin density wave correlations at this doping are short-ranged and quasistatic \cite{tranquada1997coexistence,ma2021parallel}.

Many proposals that break translational symmetry in the same way as long-range AFM---with a wavevector of $Q = \left(\pi,\pi\right)$---have been put forward, including $d$-density wave order \cite{chakravarty2001hidden}, staggered loop-current order \cite{li2019thermal}, and of course local-moment antiferromagnetism or spin-density-wave order \cite{lewin2015angle,storey2016hall}. There are also proposals that produce nodal hole pockets without breaking translational symmetry, including the Yang-Zhang-Rice (YRZ) ansatz \cite{rice2011phenomenological}, staggered fluxes \cite{wen1996theory}, and topological order \cite{scheurer2018topological}. In Supplementary Figure S2 we show that the nodal hole pockets from the YRZ ansatz also fit the ADMR data at $p = 0.21$. This suggests that the nodal hole pockets themselves, rather than the particular details of any one model, are what is important to describe the Fermi surface transformation across \ps.

Even if no static, long-range order is present in Nd-LSCO at $p =0.21$, scattering at the AFM wavevector is known to be important to many models of the pseudogap \cite{senechal2004hot,scalapino2012common,wu2017controlling}, and it may be enough for an order parameter to appear static on time scales of order of the quasiparticle lifetime ($\approx 0.1$ ps) and over length scales of order of the cyclotron radius ($\approx$ 20 nm at $B=45$~T) \cite{gannot2019fermi}. We note that there is evidence for fluctuating, short-range spin density wave correlations in Nd-LSCO near \ps \cite{tranquada1997coexistence, ma2021parallel}, and short range magnetic order has been found to onset below \ps in the related compound La$_{2-x}$Sr$_x$CuO$_4$ \cite{franchet2020hidden}. It may be that some form of this spin density wave reconstructs the Fermi surface at $p = 0.21$. Note, however that a reduction in the Hall coefficient within the pseudogap phase is universal in the cuprates \cite{badoux2016critical,collignon2017fermi,putzke2021reduced,lizaire2021transport}, and that our model of FS transformation produces the correct Hall coefficient (both above and below \ps), which strongly suggests that the model we propose here for the FS below \ps itself is universal, whereas the tendency toward spin density wave order varies substantially between different cuprates. 

Three families of unconventional superconductors---iron pnictides, organics, and heavy-fermions---share a common phase diagram in which long-range magnetic order is suppressed as a function of doping or pressure. At the critical point, where long-range order is suppressed, the superconducting \Tc is typically maximal, the resistivity is most ``anomalous'' (typically linear in temperature), and the quasiparticle mass is enhanced \cite{shishido2005drastic,walmsley2013quasiparticle,ramshaw2015quasiparticle}. Long-range magnetic order reconstructs the Fermi surface in all three classes of materials \cite{uji1997rapid,shishido2005drastic,analytis2009quantum} and thus the onset of Fermi surface transformation, near-optimal \Tc, $T$-linear resistivity, and enhanced quasiparticle interactions are tied together across dozens of superconducting materials, each with entirely different microscopic constituents. The phase diagram of the \highTc cuprates is superficially similar, with $T$-linear resistivity, near-optimal \Tc, and enhanced effective mass all occurring at a critical doping where the pseudogap phase appears. What was missing until now was direct experimental evidence of the accompanying FS transformation.

\section*{Acknowledgements}
The authors acknowledge helpful discussions with James Analytis, Debanjan Chowdhury, Nicolas Doiron-Leyraud, Nigel Hussey, Mark Kartsovnik, Steve Kivelson, Dung-Hai Lee, Sylvia Lewin, Andr\'e-Marie Tremblay, Kimberly Modic, Seth Musser, Cyril Proust, and Senthil Todadri.
A portion of this work was performed at the National High Magnetic Field Laboratory, which is supported by the National Science Foundation Cooperative Agreement No. DMR-1644779 and the State of Florida.
P.A.G. acknowledges that this project is supported by the European Research Council (ERC) under the European Union’s Horizon 2020 research and innovation program (Grant Agreement No. 681260).
J.-S.Z. was supported by an NSF grant (MRSEC DMR-1720595).
L.T. acknowledges support from the Canadian Institute for Advanced Research (CIFAR) as a Fellow and funding from the Natural Sciences and Engineering Research Council of Canada (NSERC; PIN: 123817), the Fonds de recherche du Qu\'ebec - Nature et Technologies (FRQNT), the Canada Foundation for Innovation (CFI), and a Canada Research Chair. This research was undertaken thanks in part to funding from the Canada First Research Excellence Fund. Part of this work was funded by the Gordon and Betty Moore Foundation’s EPiQS Initiative (Grant GBMF5306 to L.T.)
 B.J.R. and Y.F. acknowledge funding from the National Science Foundation under grant no. DMR-1752784.

%

\section*{Author Contributions}
A.L., P.G., L.T., and B.J.R. conceived the experiment.
J.-S.Z. grew the samples.
A.L., F.L., A.A., C.C. and M.D. performed the sample preparation and characterization.
Y.F., G.G., A.L., D.G., P.G., and B.J.R. performed the high magnetic field measurements at the National High Magnetic Field Laboratory.
Y.F., G.G., S.V., M.J.L., and B.J.R. performed the data analysis and simulations.
Y.F., G.G., S.V., P.G., L.T., and B.J.R. wrote the manuscript with input from all other co-authors.
L.T. and B.J.R. supervised the project.

\section*{Competing Interests}
The authors declare no competing interests.

\newpage
\begin{figure}[H]
\begin{center}
\includegraphics[width=.90\columnwidth]{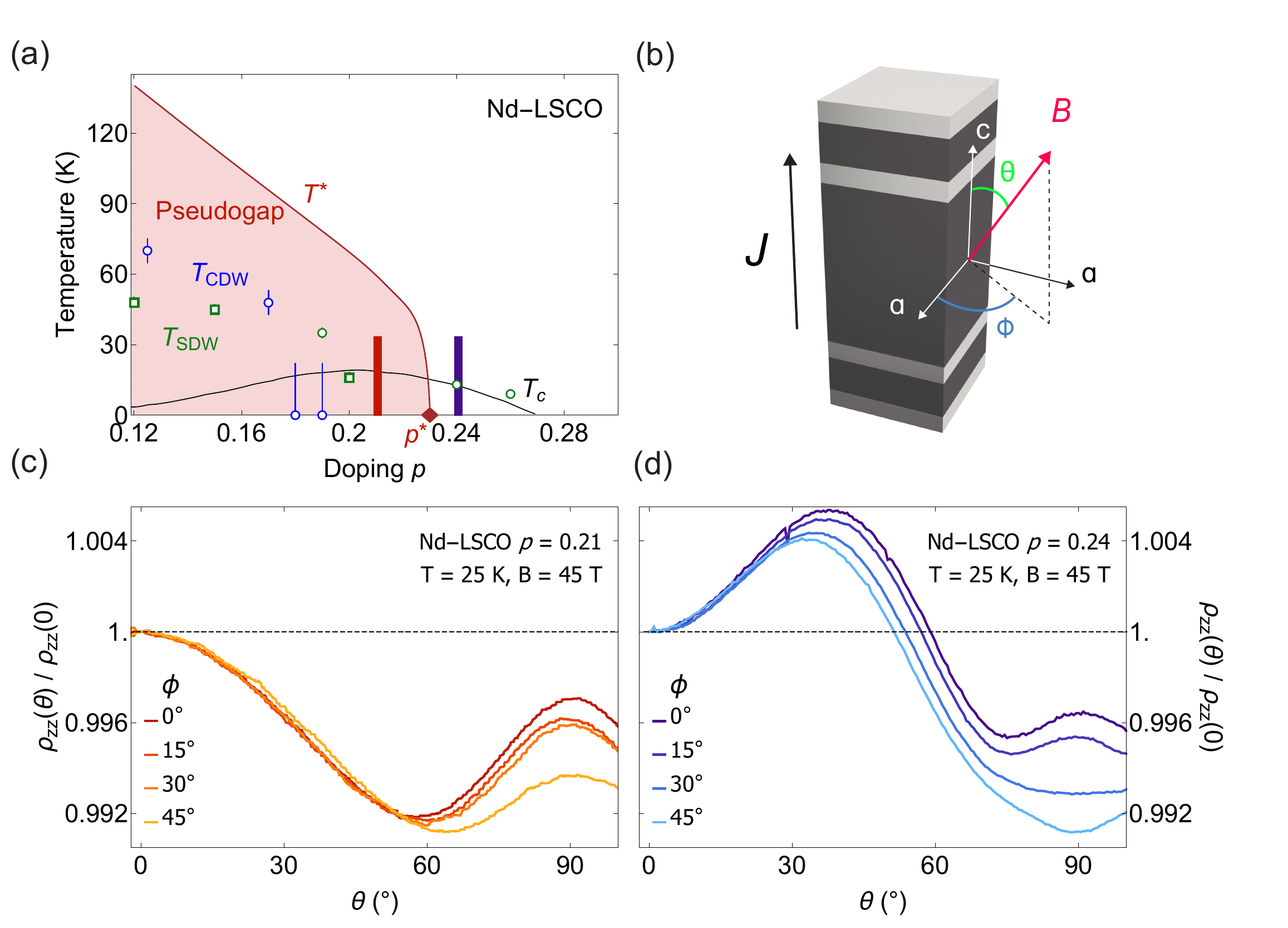}%
\end{center}
\caption{\textbf{ADMR above and below the pseudogap critical doping $\bm{p^*}$ in Nd-LSCO}.
(\textbf{a})~Temperature-doping phase diagram of the hole-doped cuprate Nd-LSCO in zero magnetic field.
The pseudogap phase is highlighted in red (the onset temperature \Ts is taken from resistivity \cite{daou2009linear, collignon2017fermi} and ARPES \cite{matt2015electron} measurements). The critical doping where the pseudogap phase ends is \ps = 0.23 (red diamond \cite{collignon2017fermi}).
The superconducting dome is marked by a solid black line and can be entirely suppressed with $\bm{B} \parallel c \ge 20$~T. The onset of short-range charge density wave order, as detected by resonant X-ray scattering \cite{gupta2020vanishing}, is indicated by the blue circles. The onset of spin density wave order, as detected by neutron scattering, is indicated with green circles (\citet{ma2021parallel}) and green squares (\citet{tranquada1995evidence}). The red and blue bars correspond to the dopings and temperature ranges measured in this study.
(\textbf{b})~Geometry of the ADMR measurements.
The sample is represented in gray with silver contacts. The black arrow identifies the direction of the electric current, $\bm{J}$, along the $c$-axis.
The angles $\phi$ and $\theta$ indicate the direction of the magnetic field $\bm{B}$ with respect to the crystallographic $a$- and $c$-axes.
(\textbf{c})~The angle-dependent $c$-axis resistivity $\rho_{\rm zz}(\theta)$ of Nd-LSCO at $p = 0.21$ ($<p^*$). All data are taken at $T=25$~K and $B=45$~T as a function of $\theta$ for $\phi=\degr{0}, \degr{15}, \degr{30},$ and $\degr{45}$, and normalized by the $\theta = 0$ value $\rho_{\rm zz}(0)$.
(\textbf{d})~Data taken under the same conditions as panel (c), but for Nd-LSCO at $p = 0.24$ ($>p^*$). Note that certain features change significantly across~\ps, including the peak near $\theta = \degr{40}$ and the $\phi$-dependence near $\theta = \degr{90}$.}%
\label{fig:phasedat}%
\end{figure}

\begin{figure}[H]
\begin{center}
\includegraphics[width=.92\columnwidth]{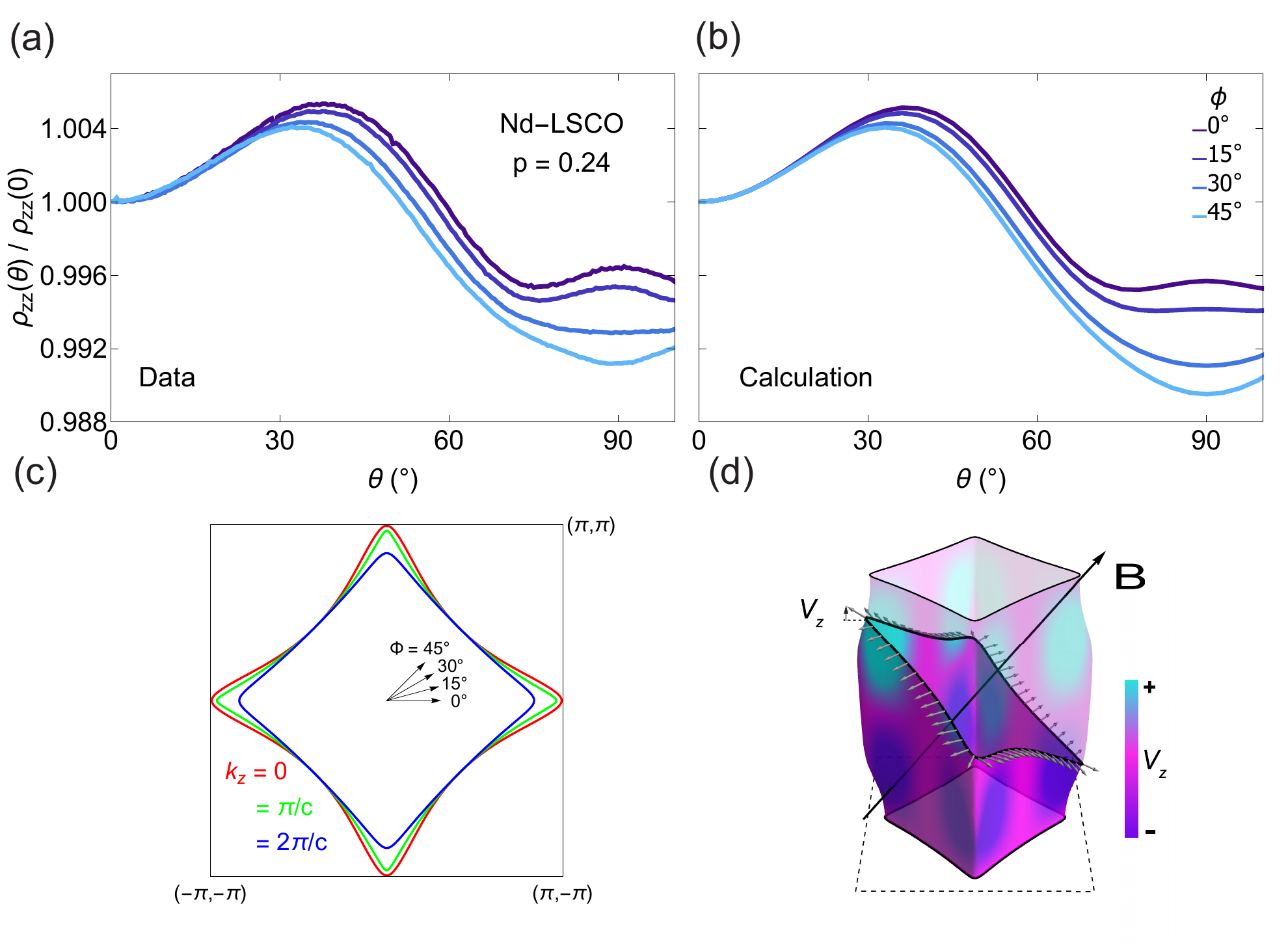}%
\end{center}
\caption{\textbf{ADMR and Fermi surface of Nd-LSCO at $\bm{p = 0.24}$.} (\textbf{a}) The ADMR of Nd-LSCO at $p = 0.24$ as a function of $\theta$ at $T=25$~K and $B=45$~T. (\textbf{b}) Simulations obtained from the Chambers formula using the tight-binding parameters from Extended Data Table 1 and the scattering rate model from \autoref{eq:scattering_cosine}.
(\textbf{c}) The Fermi surface of Nd-LSCO $p=0.24$ obtained from the ADMR calculations, with cuts shown at $k_{z}=0$, $\pi/c$, and $2\pi/c$, where $c$ is the height of the body-centered-tetragonal unit cell (and $c/2$ is the distance between copper oxide layers).
(\textbf{c}) The full 3D Fermi surface. The colouring corresponds to the $v_z$ component of the Fermi velocity, with positive $v_z$ in light blue, negative $v_z$ in purple, and $v_z=0$ in magenta. A single cyclotron orbit, perpendicular to the magnetic field, is drawn in black, with the Fermi velocity at different points around the orbit indicated with grey arrows. The strong variation in $v_z$ around the cyclotron orbit is what leads to ADMR.}
\label{fig:overdoped}%
\end{figure}

\begin{figure}[H]
\begin{center}
\includegraphics[width=.9\columnwidth]{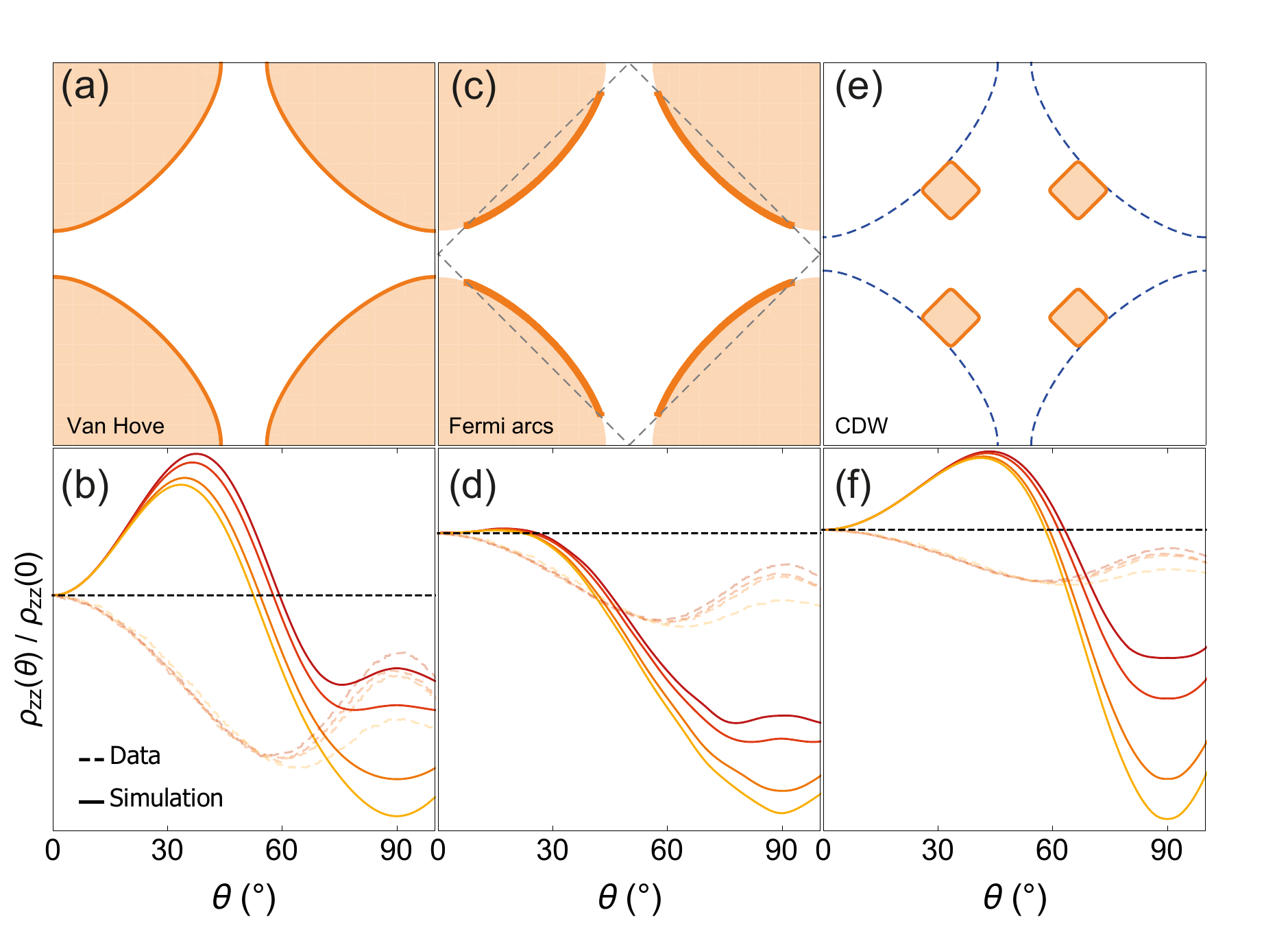}%
\end{center}
\caption{\textbf{Models that fail to account for the change in ADMR across \ps.}
The top three panels (\textbf{a, c} and \textbf{e}) show the Fermi surface for three different scenarios and the bottom three panels (\textbf{b, d} and \textbf{f}) show the resulting ADMR simulations. (\textbf{b}) ADMR calculated using the same parameters as in \autoref{fig:overdoped}a (including the scattering rate) but with the chemical potential shifted past the van Hove singularity to $p = 0.21$. The ADMR for this model is largely unchanged from the fit at $p = 0.24$. (\textbf{c}) Schematic of Fermi arcs, whereby the FS terminates at the antiferromagnetic zone boundary (grey dashed line) due to incoherence of the quasiparticles past that point. This is modelled as a scattering rate that increases considerably upon crossing the zone boundary. This model, shown in (\textbf{d}), fails to fit the data, particularly near $\theta = \degr{90}$. (\textbf{e}) Electron pockets obtained from period-3 CDW order are shown in orange, along with the original FS shown as a blue dashed line. The calculations of the ADMR for these electron pockets are shown in (\textbf{f}) but do not reproduce the data. Similar nodal electron pockets are able to account for the ADMR in YBa$_2$Cu$_3$O$_{6+x}$ at $p=0.11$ \cite{ramshaw2017broken}, where CDW order is present. }%
\label{fig:underdoped}%
\end{figure}

\begin{figure}[H]
\begin{center}
\includegraphics[width=.9\columnwidth]{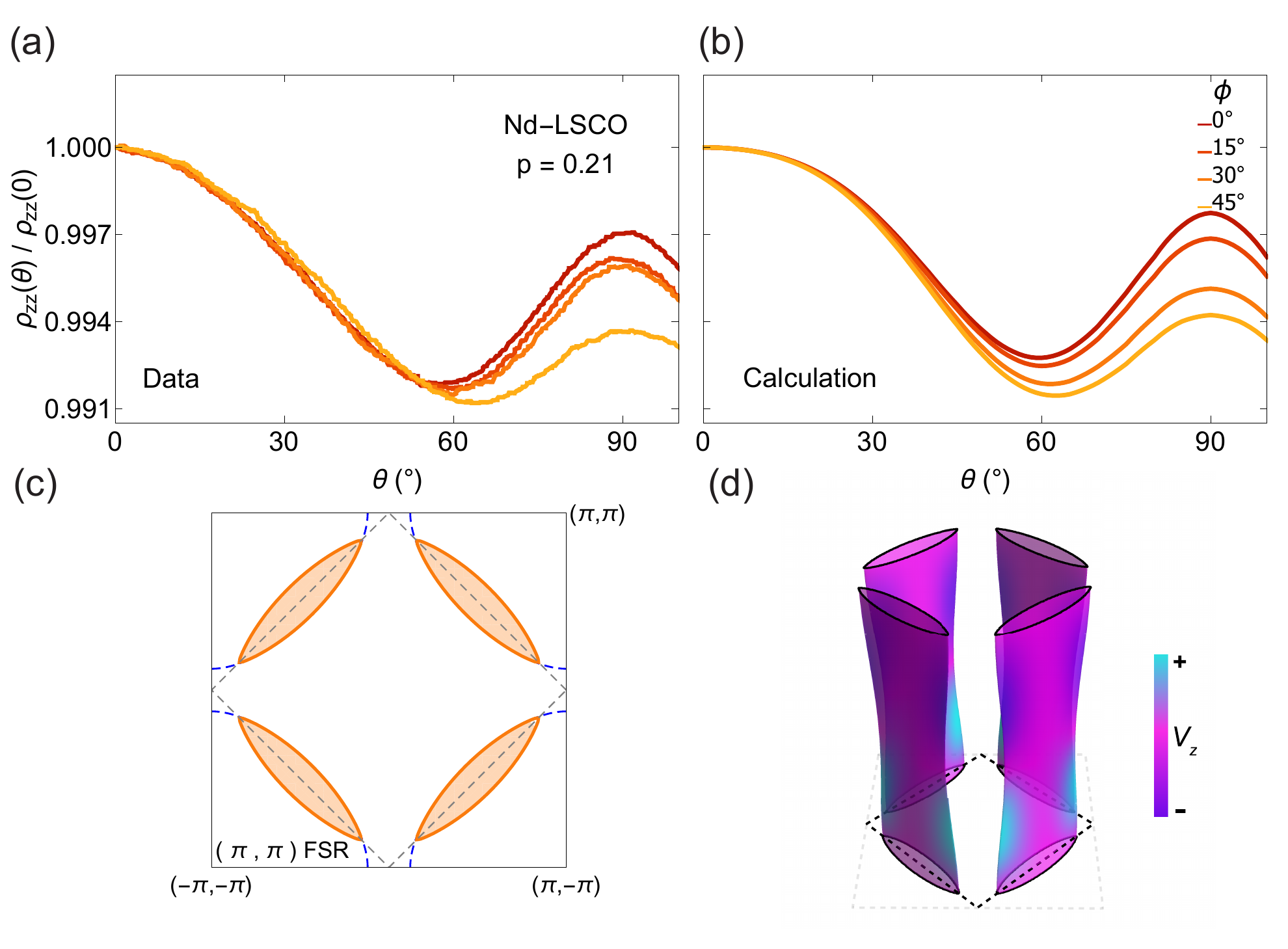}%
\end{center}
\caption{\textbf{Fermi surface reconstruction into nodal hole pockets in Nd-LSCO at $\bm{p=0.21}$.} (\textbf{a}) Measured ADMR of Nd-LSCO at $p = 0.21$ as a function of $\theta$ at $T=25$~K and $B=45$~T. (\textbf{b}) Calculated ADMR for the FS shown in (c) with an isotropic scattering rate. (\textbf{c}) FS consisting of four nodal hole pockets. These pockets are implemented via a model of antiferromagnetic order with a wavevector of $Q = \left(\pi,\pi\right)$ and a gap of 55 K, with the electron pockets removed to produce agreement with the measured Hall coefficient. (\textbf{d}) The full 3D Fermi surface at $p=0.21$ after reconstruction. }%
\label{fig:underdoped_pipi}%
\end{figure}

\newpage
%

\section*{Methods}

\renewcommand{\figurename}{Extended Data Fig.}
\newcounter{extended}
\setcounter{extended}{0}
\renewcommand\theextended{\arabic{extended}}
\renewcommand\thefigure{\arabic{extended}}

\renewcommand{\tablename}{Extended Data Table}
\newcounter{extendedt}
\setcounter{extendedt}{1}

\subsection*{Samples and Transport Measurements}
Single crystals of La$_{\rm 2-y-x}$Nd$_{\rm y}$Sr$_{\rm x}$CuO$_{\rm 4}$ (Nd-LSCO) were grown at the University of Texas at Austin using the travelling-float-zone technique, with a Nd content $y = 0.4$ and nominal Sr concentrations $x = 0.20$, $0.21$ and $0.25$. The hole concentration $p$ is given by $p = x$, with an error bar $\pm 0.003$, except for the $x = 0.25$ sample, for which the doping is $p = 0.24 \pm 0.005$ (for more details, see ref. \cite{collignon2017fermi}). The value of $T_{\rm c}$, defined as the point of zero resistance, is: $T_{\rm c} = 15.5$, $15$ and $11$~K for samples with $p = 0.20$, $0.21$ and $0.24$, respectively. The pseudogap critical point in Nd-LSCO is at $p^* = 0.23$ (ref. \cite{collignon2017fermi}).

Resistivity measurements were performed in the 45~T hybrid magnet at the National High Magnetic Field Lab in Tallahassee, USA. The sample resistance was measured with a standard 4-point contact geometry using a Stanford Research 830 Lock-In Amplifier. The samples were driven with $I_{RMS} = 1$~mA from a Keithley 6221 Current Source. Temperature was stabilized to within $\pm 1$ mK around the target temperature at each angle. Uncertainty of the absolute temperature due to thermometer magnetoresistance is negligible at $T=25$~K. The thermometer was mounted at a fixed point on the probe near the sample but not on the rotating platform. Thus, the magnetoresistance of the thermometer did not change as the sample was rotated. 

At $p = 0.21$ and $0.24$ the upper critical fields of Nd-LSCO are, respectively, 15~T and 10~T for $\bm{B} \parallel c$ \cite{michon2019thermodynamic}. By applying a magnetic field of $B=45$~T at both $T = 25$~K both samples remain in the normal state while rotating the field from $\bm{B} \parallel c$ to $\bm{B} \parallel a$.

The polar angle $\theta$ between the crystalline $c$-axis and the magnetic field was changed continuously \textit{in situ} from $\approx \degr{-15}$ to $\approx \degr{110}$  using a single-axis rotator. A voltage proportional to the angle was recorded with each angle sweep. The angle $\theta$ was calibrated by finding symmetric points in the resistivity and scaling the measured voltage such that the symmetric points lie at $\theta = \degr{0}$ and $\degr{90}$ (see Extended Data Figure 6). This procedure resulted in an uncertainty in $\theta$ of $\pm \degr{0.5}$. The azimuthal angle $\phi$ was changed by placing the sample on top of G-10 wedges machined at different angles: \degr{15}, \degr{30} and \degr{45}. An illustration of the sample mounted on the rotator stage, with a G-10 wedge to set the azimuthal angle to be \degr{30}, is shown in Extended Data Figure 6. The samples and wedges were aligned under a microscope by eye to an accuracy in $\phi$ of $\pm \degr{2}$.

\subsection*{Transport calculations in a magnetic field}
The semi-classical electrical conductivity of a metal can be calculated by solving the Boltzmann transport equation within the relaxation-time approximation. The approach most suitable for calculating angle-dependent magnetoresistance was formulated by \citet{Chambers1952kinetic}. It provides an intuitive prescription for calculating the full conductivity tensor $\sigma_{ij}$ in a magnetic field $\bm{B}$, starting from a tight-binding model of the electronic band structure~$\epsilon(\bm{k})$.
Chambers' solution is
\begin{equation}
\sigma_{ij}  = \frac{e^2}{4 \pi^3}\!\int \!d^3\bm{k} \left(-\frac{df_0}{d\epsilon}\right)v_i\!\left[\bm{k}\!\left(t=0\right)\right]\int^{0}_{-\infty}\!v_j\!\left[\bm{k}\!\left(t\right)\right]e^{t/\tau}dt,
\label{eq:chamb}
\end{equation}
where $\int \!d^3\bm{k}$ is an integral over the entire Brillouin zone, $\left(-\frac{df_0}{d\epsilon}\right)$ is the derivative with respect to energy of the equilibrium Fermi distribution function, $v_i$ is the $i^{\mathrm{th}}$ component of the quasiparticle velocity, and  $\int^{0}_{-\infty}dt$ is an integral over the lifetime, $\tau$, of a quasiparticle.
The Fermi velocity is calculated from the tight binding model as $\bm{v}_{\rm F} = \frac{1}{\hbar}\vec{\nabla}_{\bm{k}}\epsilon(\bm{k})$.
The magnetic field, including its orientation with respect to the crystal axes, enters through the Lorentz force, which acts to evolve the momentum $\bm{k}$ of the quasiparticle through $\hbar \frac{\mathrm{d}\bm{k}}{\mathrm{d}t} = e \bm{v}\times\bm{B}$. Because the magnetic field is included explicitly in this manner, Chambers' solution has the advantage of being exact to all orders in magnetic field.

The conductivity of a general electronic dispersion $\epsilon(\bm{k})$ can be calculated using \autoref{eq:chamb} \cite{goddard2004angle}. The factor $\left(-\frac{df_0}{d\epsilon}\right)$ is approximated as a delta function at the Fermi energy in the limit that the temperature $T$ is much smaller than any of the hopping parameters in $\epsilon(\bm{k})$, as is the case for our experiments. This delta function transforms the integral over the Brillouin zone into an integral over the Fermi surface, and introduces a factor of $1/|\vec{\nabla}_{\bm{k}}\epsilon(\bm{k})|$, which is the density of states. To perform the integrals in \autoref{eq:chamb} numerically, the Fermi surface is discretized, usually into 10 to 15 layers along $k_z$, with 60 to 100 points per $k_z$ layer, and each point is evolved in time using the Lorentz force equation. This moves the quasiparticles along cyclotron orbits around the Fermi surface, and their velocity is recorded at each position and integrated over time. The weighting factor $e^{t/\tau}$ accounts for the scattering of the quasiparticles as they traverse the orbit. In general, $\tau$ is taken to be a function of momentum, $\tau(\bm{k})$, and then the factor $e^{t/\tau}$ is replaced by $e^{\int_{t}^{0} dt'/\tau(\bm{k}\left(t'\right))}$. \autoref{eq:chamb} can be used to calculate any component of the semiclassical conductivity tensor.
We use it to calculate $\rho_{\rm zz}$ in \autoref{fig:phasedat}, \ref{fig:overdoped}, \ref{fig:underdoped} and \ref{fig:underdoped_pipi}. Note that, because of the highly 2D nature of the Fermi surface of Nd-LSCO, we neglect the off-diagonal components of the conductivity tensor and use $\rho_{\rm zz} \approx 1/\sigma_{\rm zz}$.

\subsection*{Nd-LSCO \textbf{\textit{p}} = 0.24: Band structure}
We use a three dimensional tight binding model of the Fermi surface that accounts for the body-centred tetragonal crystal structure of Nd-LSCO \cite{horio2018three},
\begin{equation}
  \begin{aligned}
\epsilon(k_x,k_y,k_z)=&-\mu
-2t[\cos(k_xa)+\cos(k_ya)]
\\&-4t'\cos(k_xa)\cos(k_ya)
-2t''[\cos(2k_xa)+\cos(2k_ya)]
\\&-2t_z\cos(k_xa/2)\cos(k_ya/2)\cos(k_zc/2)[\cos(k_xa)-\cos(k_ya)]^2,
  \end{aligned}
	\label{eq:band_structure_0p24}
\end{equation}
where $\mu$ is the chemical potential, $t$, $t'$, and $t''$ are the first, second, and third nearest neighbour hopping parameters, $t_z$ is the inter-layer hopping parameter, $a=3.75$~\AA~is the in-plane lattice constant of Nd-LSCO, and $c/2=6.6$~\AA~is the CuO$_2$ layer spacing. The inter-layer hopping has the form factor $\cos(k_xa/2)\cos(k_ya/2)[\cos(k_xa)-\cos(k_ya)]^2$, which accounts for the offset copper oxide planes between layers of the body-centered tetragonal structure \cite{chakravarty1993interlayer}.

The fit results are presented in \autoref{fig:overdoped}b (for the ADMR) and Extended Data Table 1 (for the tight-binding and scattering rate parameters.) Although the genetic algorithm was allowed to search over a wide range of parameters, we found that the optimal solution converged towards $t'$, $t''$ and $t_{\rm z}$ values extremely close to the ARPES values, with a 7\% deviation at most for $t_{\rm z}$. Only $\mu$, and therefore the doping $p$, is substantially different from the ARPES value. The higher doping found by ARPES may be due to the difficulty in accounting for the $k_z$ dispersion, or may be due to different doping at the surface. Nevertheless, the shape of the Fermi surface found by fitting the ADMR data (see \autoref{fig:overdoped}b, c) is electron like and qualitatively identical to the one measured by ARPES \cite{matt2015electron}, and the doping we find ($p=0.248$) is very close to the nominal one $p=0.24 \pm 0.005$ \cite{daou2009linear}.

This demonstrates that the Fermi surface is correctly mapped out by our analysis of the ADMR data. In the figures and the analysis presented in this manuscript, we use the tight-binding values from Extended Data Table 1, and for simplicity we refer to them as the ``tight-binding values from ARPES '', as they only differ by the chemical potential value.

\subsection*{Nd-LSCO \textbf{\textit{p}} = 0.24: Scattering rate model}

We have used a minimal, phenomenological, anisotropic scattering rate model to fit the ADMR data of Nd-LSCO $p=0.24$, which we refer to as the ``cosine'' model:
\begin{equation}
1/\tau(T,\phi) = 1/\tau_{\rm iso}(T)+1/\tau_{\rm aniso}(T)|\cos(2\phi)|^\nu,
\label{eq:scattering_cosine}
\end{equation}
where $1/\tau_{\rm iso}$ is the amplitude of the isotropic scattering rate, $1/\tau_{\rm aniso}$ is the amplitude of the $\phi$-dependent scattering rate, and $\nu$ is an integer. The best fit using this model is plotted in \autoref{fig:overdoped}b. The features at $\theta = \degr{40}$ and $\theta = \degr{90}$ are present with the same amplitudes as the data. With as few parameters as possible, this model captures the trend of the anti-nodal regions of the Fermi surface to have shorter quasiparticle lifetimes in the cuprates \cite{abrahams2000angle,analytis2007angle}, particularly close the van Hove singularity. This model should be seen as the simplest phenomenological model able to capture the correct shape of the real scattering rate, with the least number of free parameters. In \citet{grissonnanche2020measurement} we explore two other scattering rate models and show that they converge to the same shape as a function of $\phi$.

\subsection*{Nd-LSCO \textbf{\textit{p}} = 0.21: Large, hole like Fermi surface}
First we examine the simplest scenario to fit the ADMR at $p=0.21$, which is to keep the tight-binding values from the fit to $p=0.24$ and just shift the chemical potential across the van Hove singularity to $p=0.21$. We then explore three different types of scattering models: an isotropic scattering rate, the anisotropic ``cosine'' function from \autoref{eq:scattering_cosine}, and a model of ``Fermi arcs''. For the Fermi arcs model, the scattering rate is set to a constant value inside of the $(\pi,\pi)$ reduced Brillouin zone, and set to a different, higher value outside of this zone.

The best fit results for these three models are showed in Extended Data Figure 2b and c, and \autoref{fig:underdoped}d. They all fail to capture the correct $\phi$ dependence of the ADMR around the $\theta=\degr{90}$ feature. We conclude that the Fermi surface transformation through \ps is not the unreconstructed, hole-like Fermi surface. 

\subsection*{Nd-LSCO \textbf{\textit{p}} = 0.21: ($\pi$,$\pi$) Fermi surface reconstruction}
Several different reconstruction scenarios \cite{chakravarty2001hidden,rice2011phenomenological,storey2016hall,scheurer2018topological,li2019thermal} produce a Fermi surface that is qualitatively equivalent to the one produced by a $(\pi,\pi)$ antiferromagnetic order parameter \cite{storey2016hall}. We simulate such a reconstruction by starting with the tight binding model at ADMR values for Nd-LSCO $p=0.24$ found in Extended Data Table 1 and performing a two-dimensional $(\pi,\pi)$ reconstruction, maintaining the same interlayer coupling terms used in the unreconstructed case. The tight binding model is then
\begin{equation}
\begin{aligned}
\epsilon_{(\pi,\pi)}(k_x,k_y,k_z) =&-\mu+\tfrac{1}{2}\big[\epsilon_0(k_x,k_y,k_z) +\epsilon_0(k_x+\pi/a,k_y+\pi/a,k_z))\big]\\
&-\tfrac{1}{2}\sqrt{4\Delta^2+\big[\epsilon_0(k_x,k_y,k_z)-\epsilon_0(k_x+\pi/a,k_y+\pi/a,k_z)\big]^2}\\
&-2t_z\cos(k_zc/2)\cos(k_xa/2)\cos(k_ya/2)[\cos(k_xa-\cos(k_ya))]^2,
\end{aligned}
\label{eq:pipi}
\end{equation}
where the unreconstructed $\epsilon_0$ is given by
\begin{equation}
\begin{aligned}
\epsilon_0(k_x,k_y,k_z)&=-2t[\cos(k_xa)+\cos(k_ya)]-4t'\cos(k_xa)\cos(k_ya)\\
&-2t''[\cos(2k_xa)+\cos(2k_ya)],
\end{aligned}
\label{eq:inplane}
\end{equation}
$\Delta$ is the gap size, $t, t', t''$ represent the first, second, and third nearest neighbor hopping parameters, $\mu$ is the chemical potential, and $t_z$ is the interlayer hopping parameter.

Note that the above equations consist of a 2D antiferromagnetic model with added inter-plane hopping instead of a fully three-dimensional antiferromagnetic model. The reason for this is Nd-LSCO's tetragonal crystal structure, for which the full 3D reconstruction would induce $C_4$ rotation symmetry breaking (coming from the $[\cos(k_xa/2)\cos(k_xa/2)]$ term in the inter-plane hopping). By performing the 2D reconstruction alone, rotational symmetry in the copper-oxide planes is preserved. Moreover, such a reconstruction is likely to be more consistent with the short-length spin correlations that are incoherent between planes. Note also that short range antiferromagnetic correlations could induce a reconstruction as long as the thermal de Broglie wavelength of the electron (of order a few nanometers at 6 K given the effective mass at $p=0.21$ \cite{michon2019thermodynamic}) is shorter than the AF correlation length \cite{tremblay2012two}.

The ADMR was simulated using \autoref{eq:pipi} using a procedure similar to that described above for $p = 0.24$. It was found that an isotropic scattering rate allows to best-fit the data. Thus the scattering rate, the gap magnitude and the chemical potential were the only three parameters allowed to vary using the genetic algorithm. The best fit is presented in \autoref{fig:underdoped_pipi}f, and the fit values can be found in Extended Data Table 2.

To understand the influence of the different parameters on the fit, we show in Extended Data Figure 7 how the ADMR varies with increasing gap size. While the magnitude of the overall drop at $\theta = \degr{90}$ increases with increasing $\Delta$, the variation is rather slow and no strong qualitative change in the simulations are observed. The best fit value is found to be around $\Delta = 55$~K. We show the same but as a function of the scattering rate amplitude in Extended Data Figure 8.

The reduction in scattering rate anisotropy when moving from the unreconstructed FS at $p = 0.24$ to the nodal hole pockets at $p = 0.21$ may be due to the large reduction in the anisotropy of the density of states. Extended Data Figure 5 plots the magnitude of the Fermi velocity---inversely proportional to the density of states---for both the unreconstructed Fermi surface and the nodal hole pockets. $v_F$ varies by a factor of 25 for the unreconstructed FS, which is likely the origin of the anisotropic elastic scattering rate. At $p = 0.21$, however, $v_F$ varies by just over a factor of two---a huge reduction in anisotropy. This may explain why the scattering rate we find on the nodal hole pockets is roughly isotropic (note that scattering rate is not exactly proportional to the density of states, as it depends on the form of the scattering matrix elements.) Note that while the relative change in resistivity is reproduced by the model, the absolute value is not reproduced: the absolute resistivity at $\theta = \degr{0}$ is $\rho_{zz} = 35.80$ m$\Omega\!$ cm, whereas the fit produces $\rho_{zz} = 12.93$ m$\Omega\!$ cm. The difference between model and data may be due to incoherent contributions to the transport, which are not captured by the Boltzmann equation. 

\subsection*{Nd-LSCO \textbf{\textit{p}} = 0.21: Hall effect}

A ($\bm \pi$,$\bm \pi$) reconstruction at $p = 0.21$, with the gap value obtained by our best-fit to the ADMR data, also produces small, anti-nodal electron pockets. Although a fit can still be obtained with the electron pockets included (as their inclusion only adds more free parameters to the model), we exclude them from the model based on the calculated Hall coefficient. Extended Data Figure 4 compares the data taken at 30 K on Nd-LSCO at $p = 0.21$ (from \citet{collignon2017fermi}) with the Hall coefficient calculated from several models. Nodal hole pockets on their own produce the best agreement with the data.

\subsection*{Nd-LSCO \textbf{\textit{p}} = 0.21: CDW Fermi surface reconstruction}
A bi-axial charge density wave (CDW) with a period near to 3 lattice spacings is thought to underlie the reconstructed pocket seen in quantum oscillation experiments \cite{sebastian2014normal,ramshaw2017broken}. We simulate such a reconstruction by starting with the ARPES tight binding values for Nd-LSCO at $p=0.24$ and perform a period-three, bi-axial wavevector reconstruction of the Fermi surface. As with the ($\pi$,$\pi$) reconstruction, we perform a 2D reconstruction and maintain the same interlayer coupling terms used in the unreconstructed case. This Fermi surface reconstruction produces multiple pockets and open sheets, similar to what was shown in \citet{allais2014connecting}. We calculate the ADMR for only the diamond-shaped Fermi surface because this is the only surface that has been reported by quantum oscillations in underdoped cuprates \cite{doiron2007quantum, chan2016single}, and because it is the only Fermi surface needed to model the ADMR in YBa$_2$Cu$_3$O$_{6.6}$ \cite{ramshaw2017broken}. The inclusion of any other Fermi surfaces would lead to a value of the normal-state specific heat that is larger than what is measured \cite{riggs2011heat}.

The Hamiltonian used for finding the in-plane Fermi surface can be written as follows \cite{sachdev2013bond},
\begin{equation}
H = \sum_{\boldsymbol{k}}[\epsilon_{\rm 0}(\boldsymbol{k})c^\dagger_{\boldsymbol{k}}c_{\boldsymbol{k}}-\sum_{\boldsymbol{Q}}\Delta_{\boldsymbol{Q}}(\boldsymbol{k}+\boldsymbol{Q}/2)c^\dagger_{\boldsymbol{k}+\boldsymbol{Q}}c_{\boldsymbol{k}}],
\label{eq:cdwham}
\end{equation}
where the sum over $\bm k$ extends over the entire Brillouin zone of the square lattice, $\Delta_{\rm Q}$ is the gap of the CDW and $\bm Q$ the wave vectors of the charge ordering. For a bidirectional charge density wave with a period of three lattice spacings, the sum over $\bm Q$ extends over the 4 values $(\pm\frac{2\pi}{3},0)$ and $(0,\pm\frac{2\pi}{3})$. The in-plane electronic dispersion is the same as the in-plane dispersion $\epsilon_{\rm 0}$ described in \autoref{eq:inplane}. The Fermi surface is found by selecting the eigenvalue of the resulting $9\times9$ matrix that corresponds to the diamond-shaped Fermi surface showed in \autoref{fig:underdoped}e.

We calculate the ADMR using the Chambers formula for this model. The result is shown in Extended Data Figure 3 for a number of different CDW strengths and a $d-$wave form factor. The simulated ADMR is somewhat reminiscent of the $p = 0.24$ data, except that the peak that was found at around $\theta=\degr{30}$ for $p=0.24$ has been pushed out to $\theta = \degr{60}$. This qualitative similarities arise because both the unreconstructed Fermi surface and the small reconstructed diamond are similar in shape. The features are pushed to higher $\theta$ for the reconstructed case because $k_F$ is smaller. It is clear, however, that a CDW reconstruction does not match the ADMR for Nd-LSCO~$p=0.21$.

\makeatletter
\apptocmd{\thebibliography}{\global\c@NAT@ctr 57\relax}{}{}
\makeatother

\renewcommand\refname{Methods References}

\section*{Data Availability}
Experimental data presented in this paper is available at http://wrap.warwick.ac.uk/161600/. The results of the conductivity simulations are available from the corresponding authors upon reasonable request.

 \section*{Code Availability}
The code used to compute the conductivity is available from the corresponding authors upon reasonable request.

\newpage

\begin{table}[h!]
\scriptsize
  \begin{center}
    \begin{tabular}{|c c c c c c c c c c|}
    \hline
     & $t$ (meV) & $t'$ & $t''$ & $t_z$ & $\mu$ & $p$& $1/\tau_{\rm iso}$ (ps$^{-1}$) & $1/\tau_{\rm aniso}$ (ps$^{-1}$) & $\nu$\\
     ADMR  & $160\pm 30$ & $-0.1364t$ & $0.0682t$ & $0.0651t$ & $-0.8243t$ & 0.248 &$12.595 \pm 0.002$ & $63.823 \pm 0.26$ & $12 \pm 1$\\
    ARPES & 190 & -0.136t & 0.068t & 0.07t &  & 0.28 & & & \\

    \hline
    \end{tabular}
  \end{center}
	\refstepcounter{extendedt}\label{tab:fit_result_0p24_bandstructure}
 \caption{\textbf{Tight-binding parameters from the fit to the ADMR~data at $\bm{p}$ = 0.24.} Best fit tight-binding values for the Nd-LSCO $p=0.24$ ADMR data (using the cosine scattering rate model of \autoref{eq:scattering_cosine}). The results are extremely close to ARPES tight-binding values reported in \citet{matt2015electron} and \citet{horio2018three}, reproduced here on the second line. Error bars on the AMDR-derived hopping parameters and chemical potential are all $\pm0.0005$, and were obtained following the procedure described in the above section. The error bar on the value of $t_z$ measured by ARPES is $\pm 0.02t$ (J. Chang and M. Horio, private communication.)}
\end{table}

\begin{table}[h!]
\scriptsize
  \begin{center}
    \begin{tabular}{|c c c c|}
    \hline
    $T$ (K)& $\mu$ & $1/\tau_{\rm iso}$ (ps$^{-1}$) & $\Delta$ (K)\\
 
    25 & $-0.495t\pm0.01$ & $22.88\pm0.30$ & $55\pm11$\\
    \hline
    \end{tabular}
  \end{center}
		\refstepcounter{extendedt}
 \caption{\textbf{Results of the fit of the Nd-LSCO $\bm{p}$ = 0.21 data  with ($\bm \pi$,$\bm \pi$) reconstruction.} Fit parameter values for Nd-LSCO $p=0.21$ plotted in \autoref{fig:underdoped_pipi}f. The band structure parameters were kept fixed at ARPES values \cite{matt2015electron}. Error bars were obtained following the procedure described in the above section.}
    \label{tab:fit_result_0p21}
\end{table}

\newpage

\begin{figure}[h]
\begin{center}
\includegraphics[width=0.7\columnwidth]{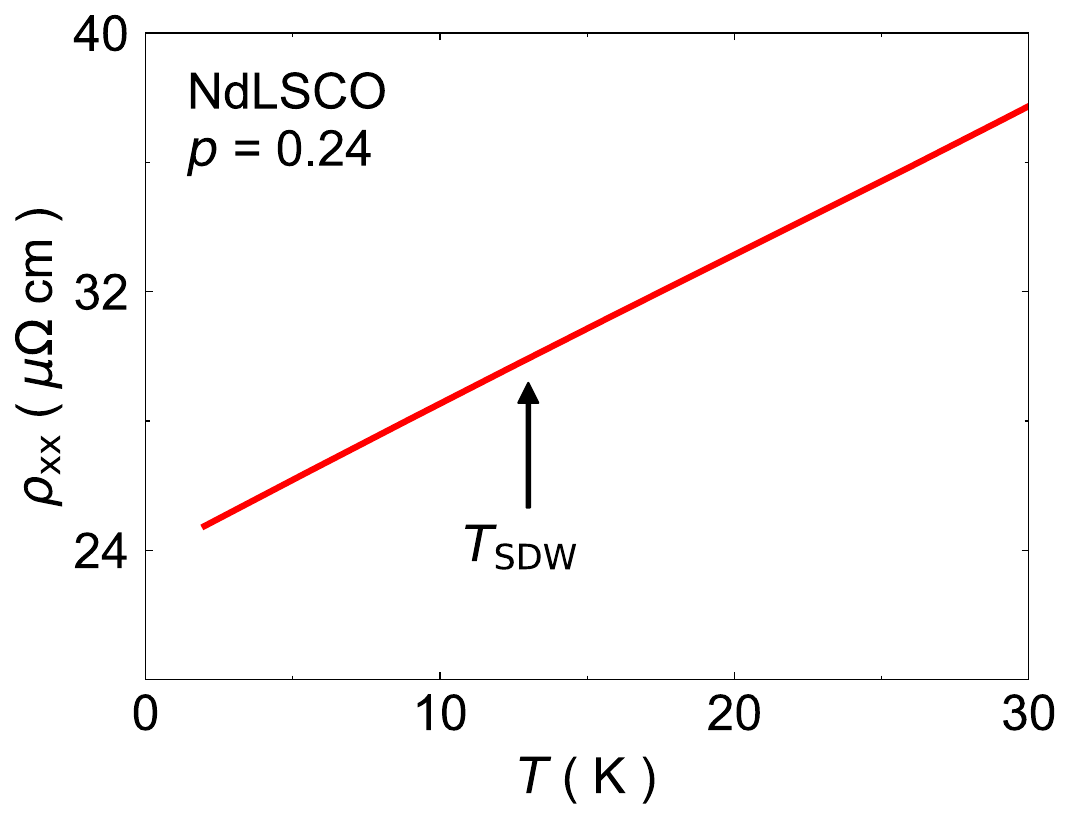}
\end{center}
\refstepcounter{extended}
\caption{\textbf{Resistivity of Nd-LSCO at $\bm p$~=~0.24 near $T_{\rm SDW}$ as determined by \citet{ma2021parallel}.} In-plane resistivity data at $B$~=~35~T as a function of temperature. The resistivity $\rho_{\rm xx}$ (red line) is perfectly linear down to the lowest temperature without any sign of an upturn or even a change in slope at $T_{\rm SDW} = 13 \pm 1$~K (black arrow) reported by \citet{ma2021parallel} at $B~=~0$~T. This suggests that either the SDW is not present in our samples or that the SDW vanishes in a magnetic field and thus does interfere with our measurements performed at $B$~=~45~T.}
\label{fig:neutron_vs_rhoxx}
\end{figure}

\begin{figure}
\begin{center}
\includegraphics[width=1\columnwidth]{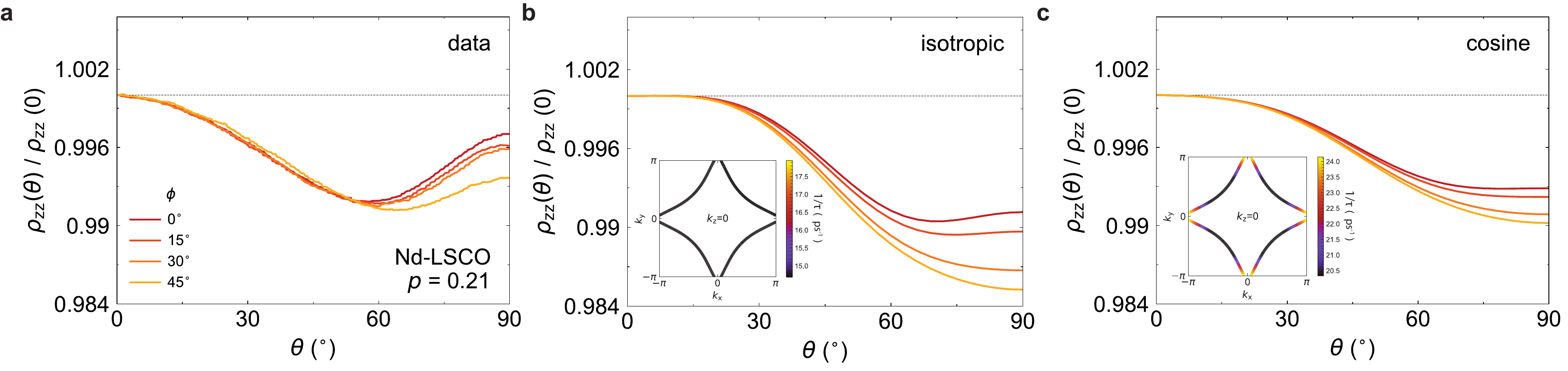}
\end{center}
\refstepcounter{extended}
\caption{\textbf{Best fit of Nd-LSCO $\bm{p}$~=~0.21 data with the large, hole-like, unreconstructed Fermi surface.}  (\textbf{a}) ADMR data on Nd-LSCO $p=0.21$ at $T=25$~K and $B=45$~T; (\textbf{b}, \textbf{c}) The best fits for the ADMR data in (a) using the band structure ARPES values for Nd-LSCO $p=0.24$ with the chemical potential shifted across the van Hove point (at $p\approx 0.23$) to $p=0.21$, where the Fermi surface is hole-like. Insets represent the scattering rate distribution values over the hole-like Fermi surface at $p=0.21$. In (b), the scattering is isotropic over the Fermi surface; in (c) we use the cosine scattering rate model (this figure differs from \autoref{fig:underdoped}b because there we only shift the chemical potential, while here we show the best-fit using this model.)}
\label{fig:scattering_models_0p21_large_hole}
\end{figure}

\begin{figure}[h]
\begin{center}
\includegraphics[width=\columnwidth]{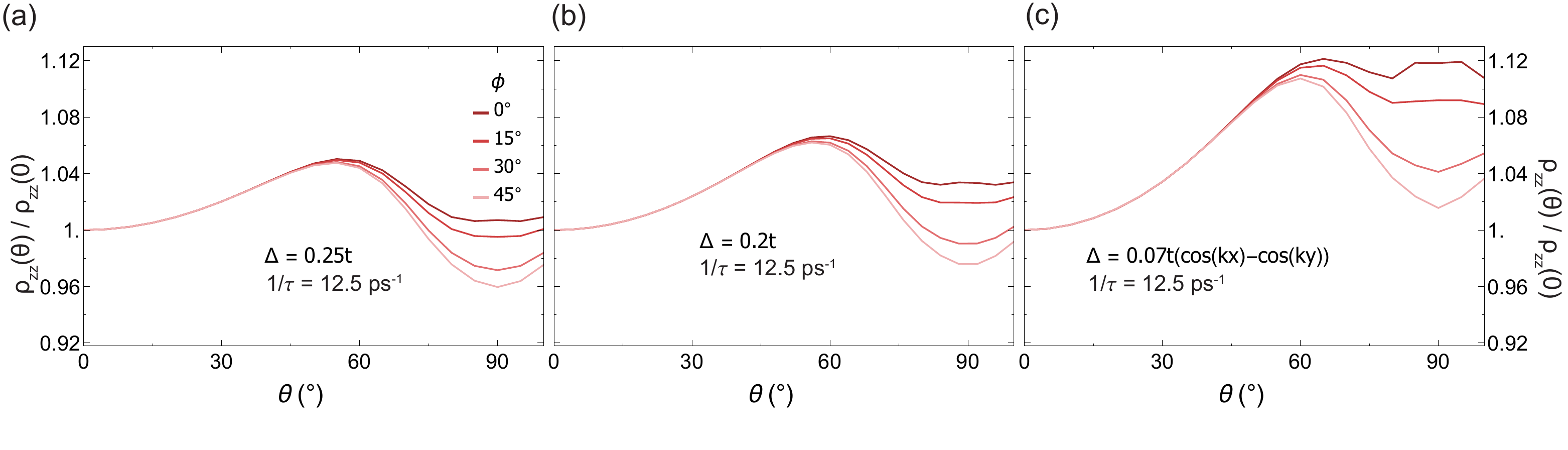}%
\end{center}
\refstepcounter{extended}
\caption{\textbf{Calculation of ADMR for a period three CDW Fermi surface reconstruction.} Calculations using two different gap sizes are shown in (a) and (b), and using a $d-$wave form factor is shown in (c).}%
\label{fig:cdw_reconstruction}%
\end{figure}

\begin{figure}[h]
\begin{center}
\includegraphics[width=.65\columnwidth]{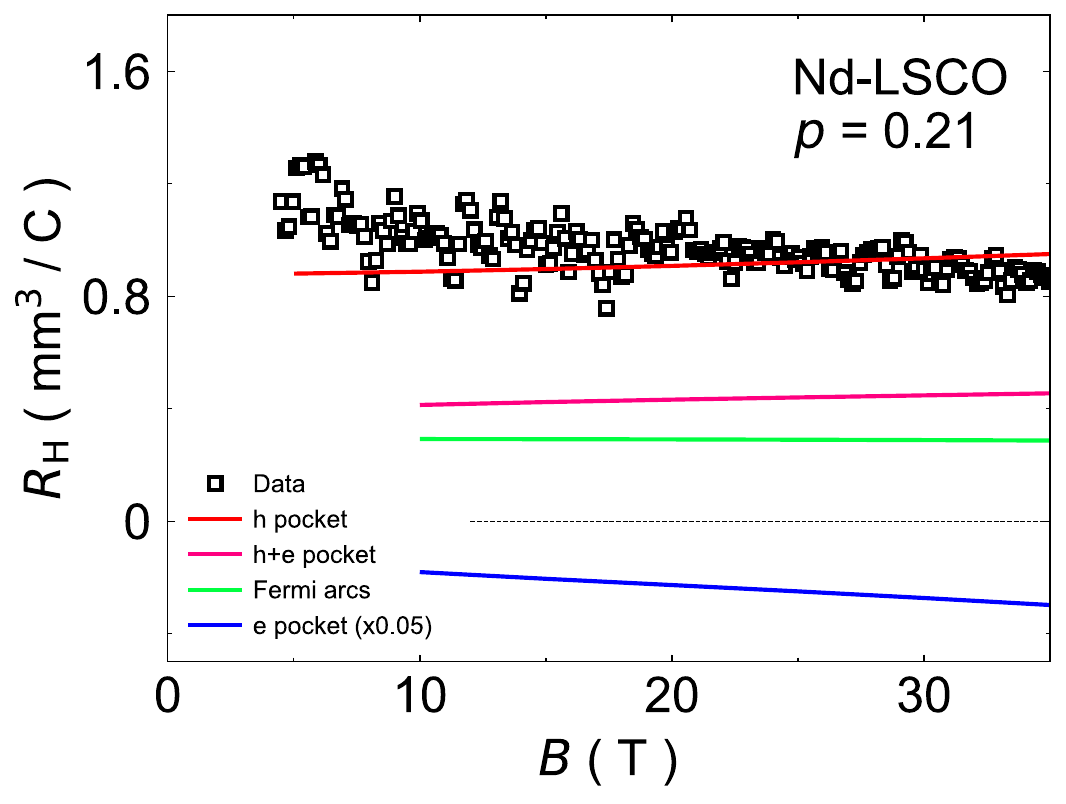}
\end{center}
\refstepcounter{extended}
\caption{\textbf{The Hall effect in Nd-LSCO at $p = 0.21$.} The data is taken at 30 K and is reproduced from \citet{collignon2017fermi}. ``h pocket'' is from the fit to the data shown in Figure 4 of the main text; ``h+e pocket'' is from a fit that includes both the hole and electron pockets after ($\bm \pi$,$\bm \pi$) reconstruction; ``Fermi arcs'' is from the fit in Figure 3c and d of the main text; ``e pocket'' is from just the electron pocket produced by ($\bm \pi$,$\bm \pi$) reconstruction, scaled down by a factor of 20 for clarity.  }
\label{fig:hall}
\end{figure}

\begin{figure}[h!]
\begin{center}
\includegraphics[width=.5\columnwidth]{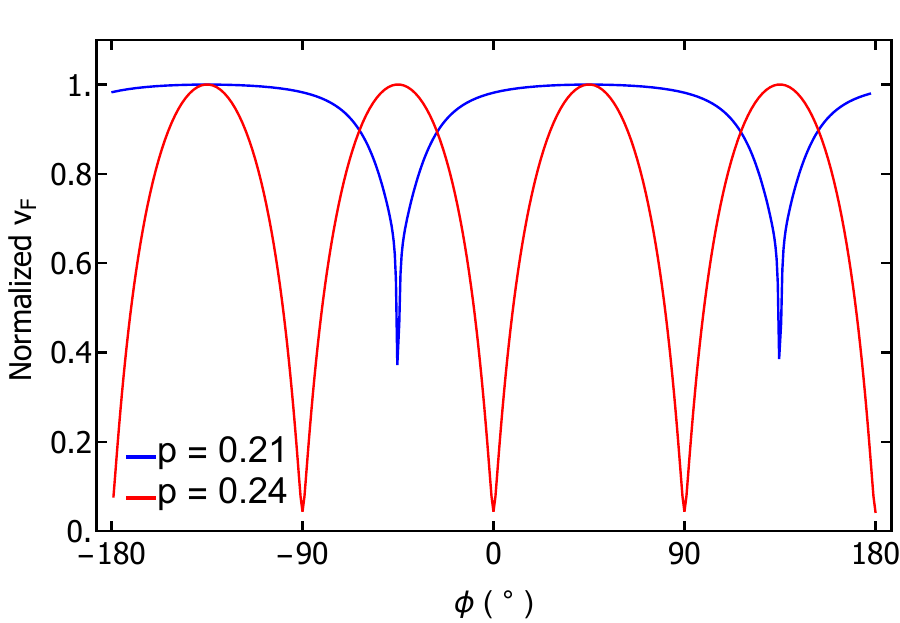}
\end{center}
\refstepcounter{extended}
\caption{\textbf{Variation in the Fermi velocity around the Fermi surface above and below \ps.} The red curve plots the magnitude of the Fermi velocity around the Fermi surface at $p = 0 .24$, as shown in \autoref{fig:overdoped}. The blue curve plots the same quantity for a single nodal hole pocket, as shown in \autoref{fig:underdoped_pipi} (the reduction in symmetry is because each nodal hole pocket is 2-fold symmetric). The total anisotropy in $v_F$ around the Fermi surface is a factor of 25 at $p = 0.24$, but just larger than a factor of 2 at $p = 0.21$.}
\label{fig:vf}
\end{figure}

\begin{figure}%
\begin{center}
\includegraphics[width=1\columnwidth]{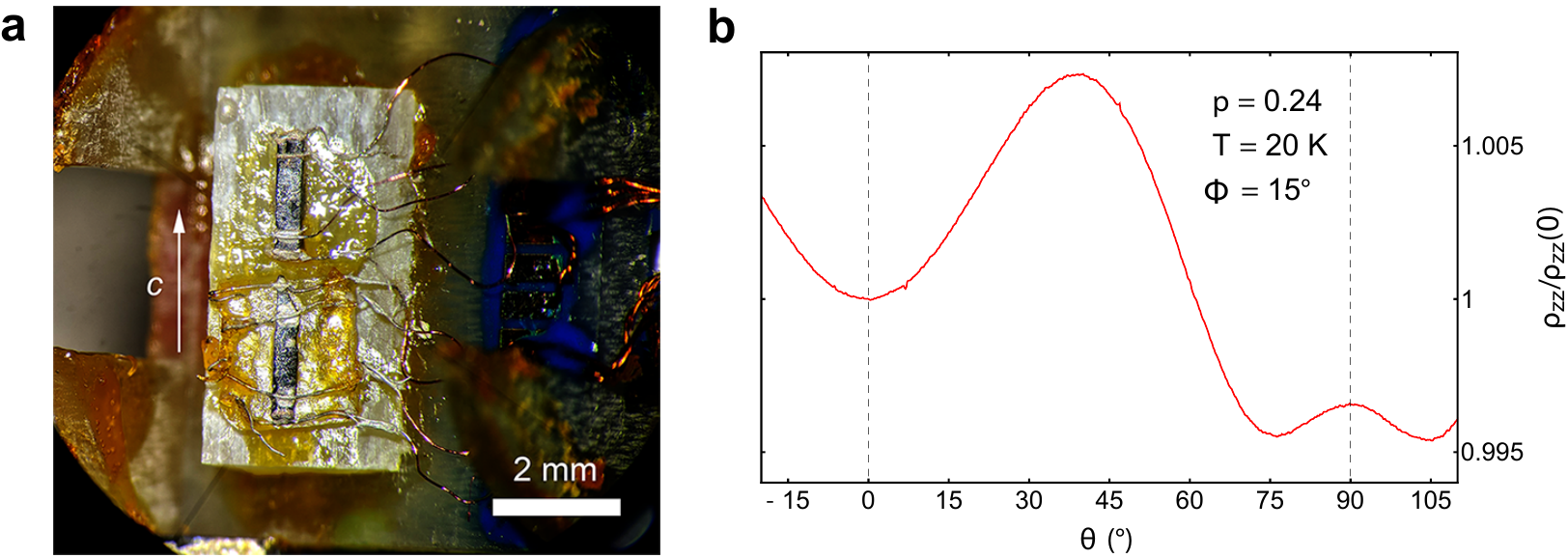}
\end{center}
\refstepcounter{extended}
\caption{\textbf{ADMR experimental set up.} (\textbf{a}) An illustration of the sample mounting. The two samples here are mounted on a G-10 wedge to provide a $\phi$ angle of \degr{30}. Additional wedges provided angles of $\phi = \degr{15}$ and $\degr{45}$; (\textbf{b})~ADMR as a function of $\theta$ angle from \degr{-15} to \degr{110} and $\phi=0$ at $T=20$~K for Nd-LSCO $p=0.24$, showing the symmetry of the data about these two angles.}
\label{fig:sampleMount}
\end{figure}

\begin{figure}[h!]
\begin{center}
\includegraphics[width=\columnwidth]{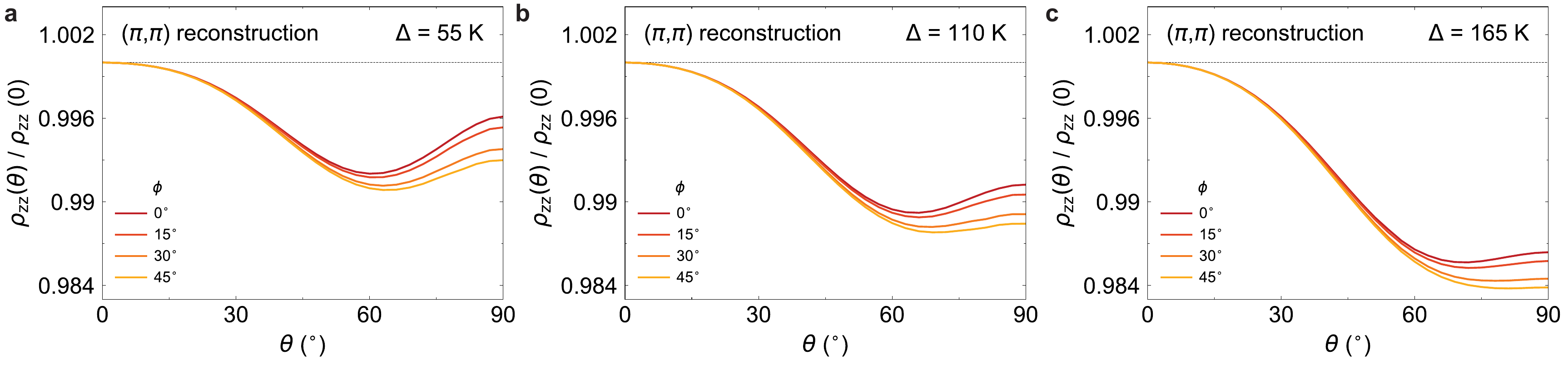}
\end{center}
\refstepcounter{extended}
\caption{\textbf{ADMR dependence on the gap amplitude with ($\bm \pi$,$\bm \pi$) reconstruction.} ADMR calculations with a ($\pi$,$\pi$) reconstructed Fermi surface for different gap amplitudes at fixed isotropic scattering rate value $1/\tau=22.88$~ps$^{-1}$. Note that this within $\approx 40$\% of the nodal scattering rate at $p=0.24$, consistent with a nodal hole pockets reconstructed from a the larger Fermi surface.}
\label{fig:pipi_gap_evolution}
\end{figure}

\begin{figure}[h!]
\begin{center}
\includegraphics[width=\columnwidth]{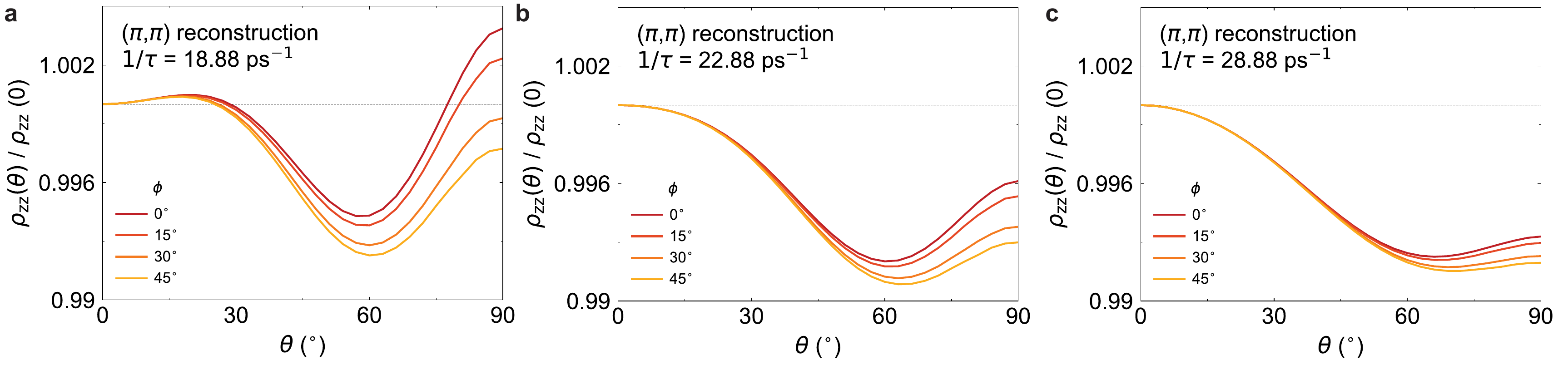}
\end{center}
\refstepcounter{extended}
\caption{\textbf{ADMR dependence on the scattering rate amplitude with ($\bm \pi$,$\bm \pi$) reconstruction.} ADMR calculations with a ($\pi$,$\pi$) reconstructed Fermi surface for different isotropic scattering rate amplitudes at fixed gap value at $\Delta=55$~K.}
\label{fig:pipi_scattering_evolution}
\end{figure}

\end{document}